\documentstyle[epsf,12pt]{article}

\def\Dirac#1{#1\hskip-6pt/}
\def\dslash{\Dirac\partial}
\def\smhalf{{\textstyle{\frac{1}{2}}}}
\def\smquarter{{\textstyle{\frac{1}{4}}}}
\def\fpj{\hspace{-7mm}}

\begin{document}

\title{The Nonperturbative Color Meissner Effect in a Two-Flavor Color
Superconductor} 

\author{ 
G. W. Carter\footnote{E-mail: carter@tonic.physics.sunysb.edu} \\
  {\small {\it Department of Physics and Astronomy,}}\\
  {\small {\it SUNY Stony Brook, NY, 11794-3800 USA}}\\
D. Diakonov\footnote{E-mail: diakonov@nordita.dk} \\
  {\small{\it NORDITA, Blegdamsvej 17, DK--2100 Copenhagen \O, Denmark}} }

\date{\today}
\maketitle

\begin{abstract}

Color superconductivity in QCD
breaks the $SU(3)$ color gauge group down to $SU(2)$, inducing masses in
five of the eight gluons. This is a dynamical Higgs effect, in which the
diquark condensate acts as the vacuum expectation value of a composite
scalar field. In order to analyze this effect at low quark density, when
gaps are large and generated nonperturbatively, we use instanton-induced
quark interactions augmented with gauge-invariant interactions between
quarks and perturbative gluons. The five gluon masses are found from the
static limit of the relevant polarization operators, in which
transversality is maintained via the Nambu-Goldstone modes of broken color
symmetry. Working in the microscopic theory we calculate these masses to
one-loop order and estimate their density dependence. Finally, we
speculate that the Meissner effect may postpone the onset of color
superconductivity to higher matter density than estimated previously.

\end{abstract}

\section{Introduction}

It has recently been demonstrated that the color symmetry of Quantum
Chromodynamics (QCD) might be spontaneously broken through the formation
of a diquark condensate.  This phase is metastable at zero and small
matter density \cite{DFL} but becomes stable and replaces the usual
chiral-broken phase at some critical density \cite{ARW,RSSV}.

Condensation of diquarks is analogous to Cooper pairing of
electrons in BCS theory, in which the massive photon is
the microscopic embodiment of the Meissner effect. In QCD the gauge
group is $SU(3)$, but the consequences are similar: gauge bosons
(gluons) become massive due to their interaction with the correlated
fermions (quarks). Not every gluon species becomes massive, since the
symmetry breaking pattern is $SU(3) \rightarrow SU(2)$ and the
three gluons of the residual $SU(2)$ remain massless. Among the five
massive modes, four are degenerate and the mass of the fifth is
$\sqrt{4/3}$ times that of the others. This is comparable to the
symmetry-broken phase of the electroweak theory. Thus the direct analogy
is to the Higgs mechanism, with the diquark field forming a composite
scalar and its nonzero VEV generating a dynamical Higgs effect.

In this paper this mechanism is analyzed microscopically.
As the source of diquark condensation
we will use the effective quark action obtained by averaging over
instanton configurations \cite{DD}.  In this approach to the QCD vacuum,
all information from the large background gauge fields (instantons) is
encoded in the non-local form of the 't Hooft interaction and the two
instanton parameters, average size ($\rho$) and number density ($N/V$).
This procedure replaces dynamical gluons with
classical, nonperturbative field solutions, ignoring quantum gluonic
fluctuations.
The reintroduction of gluonic fluctuations will require a
gauge-invariant perturbative modification of the 't Hooft vertex. A
direct consequence of gauge invariance is the transverse polarization
operator, which we will compute explicitly. This is the main result of
the paper as from it we obtain the gluon masses.

Since the dynamical Higgs effect is interesting regardless of density, we
will first consider the vacuum alone (chemical potential $\mu=0$),
although in this case the superconductor is only metastable \cite{DFL}.
The gluon masses are found to be proportional to $g\Delta$ where $\Delta$ is
the superconducting gap, the scale of which is a nonperturbative one set
by the strength of the instanton background.  Extending the formalism to
finite chemical potential $\mu$ is straightforward, given that the density
dependence of the propagators and effective action are known \cite{CD}.
However, in this paper we avoid exact calculations at $\mu\neq 0$.
Instead, for simplicity we assume that the in-medium gluon masses are
determined by the behavior of $g\Delta(\mu)$ up to and somewhat above the
critical value of $\mu\approx 300$ MeV. At this density, the point at
which the color superconductor becomes the stable phase, we find that the
gluon masses are about 120 MeV.

For gluons the color-breaking Meissner masses are a primary
ramification of the superconducting state, however this is not
the only effect of a quark medium. There will be additional
contributions, among them a Debye screening mass of the order of
$g\mu$, which do not break color symmetry. When the density becomes
large, the Debye mass increases and instantons are screened out of the
picture.  Yet diquark condensation persists, now due to perturbative
gluon exchange \cite{BL}, and at asymptotically large density the
Meissner masses are also proportional to $g\mu$ \cite{SS1,DR}.  So while
one still has quark pairing, any matching between low and high density
mechanisms remains unclear.

In this paper we consider the case of two massless flavors, which is
expected to be relevant at finite-density chiral restoration given a
relatively large strange quark mass \cite{ABR1,SW}.
Instantons are Euclidean pseudoparticles, and thus all calculations will
be in Euclidean space.

In Section 2 we recall those features of the ordinary Higgs mechanism
(based on elementary scalar fields) which will also be relevant for
composite scalars. In Section 3 the instanton-induced action including
perturbative gluons will be formulated. In Section 4 the diquark gap
equation is reviewed, and in Section 5 the color current is determined.
In Section 6 we sum a set of quark-quark correlation functions to recover
the Nambu-Goldstone modes of the theory, a necessary exercise as these
will mix with the longitudinal gluons.  These ingredients are assembled
into the gluon polarization operator in Section 7, which taken in the
static limit corresponds to a mass. 
Section 8 compares the result to chiral symmetry breaking, and
in Section 9 this result is discussed and conclusions are drawn.

\section{Higgs Mechanism}

The dynamical Higgs mechanism, though technically more involved,
does not differ much from the ordinary one. Let us denote the
quark bilinear combination
\begin{equation}
\phi^\alpha = \epsilon_{\alpha\beta\gamma}\epsilon_{fg}\epsilon_{ij}\;
\left(\psi_L^{\beta f i}\,\psi_L^{\gamma g j} + (L\to R)\right),
\label{scalar} \end{equation} where Greek letters denote color, $f,g=1,2$
flavor, and $i,j=1,2$ are spinor indices of the left ($L$) and right ($R$)
components of the quark field $\psi$. The resulting complex field
$\phi^\alpha$ is a Lorentz scalar isoscalar field belonging to the
fundamental representation of the $SU(3)$ group. To support gauge
invariance it must couple to the gauge potential via the covariant
derivative, $(\nabla_\mu)^\alpha_\beta=\partial_\mu\delta^\alpha_\beta- ig
A_\mu^a(\lambda^a/2)^\alpha_\beta$, in which the $\lambda^a$ denote the
eight Gell-Mann matrices. The kinetic energy term for the composite
diquark field $\phi^\alpha$ is the usual 
\begin{eqnarray} 
{\cal L}&=&Z_\phi^{-1}\left|\nabla_\mu \phi\right|^2 \nonumber\\
&=&Z_\phi^{-1}\Bigg[\partial_\mu\phi^\dagger_\alpha\partial_\mu\phi^\alpha
+i\frac{g}{2}A_\mu^a\left(\phi^\dagger_\alpha(\lambda^a)^\alpha_\beta
\partial_\mu\phi^\beta-\partial_\mu\phi^\dagger_\alpha
(\lambda^a)^\alpha_\beta\phi^\beta\right) 
 \nonumber\\ && 
+\frac{g^2}{4}A_\mu^aA_\mu^b\phi^\dagger_\alpha
(\lambda^a\lambda^b)^\alpha_\beta\phi^\beta\Bigg]. 
\label{lagr}
\end{eqnarray} 
The only difference with the standard case of the
elementary field is that there is, in principle, a common `wave function
renormalization' factor, $Z_\phi$. For elementary fields $Z_\phi=1$ at the
tree level, however, it deviates from unity even for the elementary field
when one takes into account the perturbative virtual emission of
particles. For the fields that are composite from the start there is no
reason for $Z_\phi$ to be unity. This quantity is {\it a priori} unknown
and should be determined from a dynamical calculation, as will follow below. We
stress, however, that the relative weights of the three terms in
Eq.\,(\ref{lagr}) are fixed by gauge invariance.

If the scalar field $\phi^\alpha$ develops a nonzero VEV signaling
the diquark condensation,
\begin{equation}
\langle\phi^\alpha\rangle=2\Delta_0\,\delta^{\alpha 3},
\label{diqcond}
\end{equation}
(it can be always arranged along the third color axis and made real), then
gluons obtain a mass matrix 
\begin{equation}
M^2_{ab}=2g^2Z_\phi^{-1}\Delta_0^2\,(\lambda^a\lambda^b)^3_3
=\left\{\begin{array}{ll} 0&\qquad a=b=1,2,3\\
2g^2Z_\phi^{-1}\Delta_0^2&\qquad a=b=4,5,6,7\\
\frac{8}{3}g^2Z_\phi^{-1}\Delta_0^2&\qquad a=b=8. \end{array}\right.
\label{massmatr} 
\end{equation} 
The symmetry breaking pattern is, thus,
$SU(3)\to SU(2)$; three gluons corresponding to the unbroken $SU(2)$
subgroup remain massless, four gluons obtain masses proportional the Higgs
VEV, and the fifth gluon is $\sqrt{4/3}$ times as heavy. This relation will
be, of course, reproduced for composite Higgs fields as well, as it
follows from symmetry considerations alone.

\begin{figure}[bt]
\setlength\epsfxsize{6.2cm} 
\centerline{\epsfbox{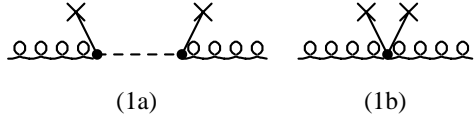}}
\caption{Diagrams contributing to the gluon polarization operator in the
Higgs effective theory.} 
\label{hpol} 
\end{figure} 
This simple elementary-Higgs model also provides an alternative way to find the
gluon masses, which we will generalize to the case of the composite Higgs
field. One can compute the gluon polarization operator
$\Pi_{\mu\nu}^{ab}(q)$, as seen in Fig.~\ref{hpol}. In Fig.~\ref{hpol}a
we take the linear coupling of the gauge potential $A_\mu^a$, as given
by the second term in Eq.\,(\ref{lagr}), and iterate it twice. The
intermediate state in this diagram is the would-be Nambu-Goldstone
boson, which contributes to the polarization operator
\begin{eqnarray}
\Pi_{\mu\nu}^{({\rm NG})\,ab}(q)
&=&2\left[Z_\phi^{-1}g\langle\phi^\dagger_\alpha\rangle
(\lambda^a)^\alpha_\gamma q_\mu\right]\;
\left[\frac{Z_\phi}{q^2}\right]\;
\left[q_\nu\,Z_\phi^{-1}g(\lambda^b)^\gamma_\beta
\langle\phi^\beta\rangle\right]
\nonumber\\
&=&2g^2Z_\phi^{-1}\Delta_0^2\;\frac{q_\mu q_\nu}{q^2}
(\lambda^a\lambda^b)^3_3.
\label{Pia}\end{eqnarray}
This contribution is purely longitudinal.

Fig.~\ref{hpol}b is the contact term; it gives a Kronecker delta
contribution:
\begin{equation}
\Pi_{\mu\nu}^{({\rm contact})\,ab}(q)
=-2 g^2 Z_\phi^{-1}\Delta_0^2\;\delta_{\mu\nu}(\lambda^a\lambda^b)^3_3.
\label{Pib} \end{equation} Combining the two we get a transverse
polarization operator, 
\begin{equation} 
\Pi_{\mu\nu}^{({\rm full})\,ab}(q) =-2g^2 Z_\phi^{-1}\Delta_0^2\;
\left(\delta_{\mu\nu} -\frac{q_\mu q_\nu}{q^2}\right)(\lambda^a\lambda^b)^3_3. 
\label{Pi} 
\end{equation} 
The fact that it is transverse reflects the color current conservation, even
in case of a broken symmetry.

In the case of a composite Higgs field the tree diagrams of
Fig.~\ref{hpol} will be replaced by loop diagrams, formed by quarks,
however, the general setting will be rather similar to this simple case:
there will be a pole contribution from the would-be Goldstone bosons in
the intermediate state, and there will be a `contact' term. The resulting
polarization operator will be transverse, {\em provided} one takes a
conserved color current.

\section{Gauge Invariant Effective Action}

Following a long tradition of instanton phenomenology, we assume
the following:
\begin{itemize}
\item Instantons are the dominant nonperturbative contribution to
low-energy QCD, and
\item
Low and high momentum scales are safely separable for quark and gluon
fields. \end{itemize} These assumptions are supported {\em a posteriori}
by instanton-based phenomenology of the vacuum, as reviewed in
Ref.~\cite{SS2}.  Instantons break chiral symmetry spontaneously and axial
$U(1)$ anomalously, both at a satisfactory magnitude. Instanton-induced
interactions also provide the necessary $qq$ attraction for a diquark
condensate, nonperturbatively.

Here we use a formulation which not only relies on the separation of
quark zero and free modes, but also explicitly includes both instantons
and perturbative gluons. Gluons are separated into classical instantons
and quantum corrections, which we write as 
\begin{equation} 
A_{\mu}(x) = \sum A_{\mu}^{(I)}(x) + \sum A_{\mu}^{(\bar I)}(x) + a_{\mu}(x),
\end{equation} 
where the sums are over all instantons and anti-instantons,
each given by the 't Hooft solution in the singular gauge \cite{tH}.
Although there is a certain distribution of instanton sizes
(see, for example, \cite{latsize})
we simply use the average value, $\rho \simeq 1/3$ fm, in all
calculations.

The main idea of this paper is to examine the effects of color
superconductivity on the gluonic excitations above the instanton
vacuum, which means we must retain and analyze the $a_{\mu}^a$.
Hereafter, the term `gluon' will refer to the perturbative gauge
fluctuation above the instanton background, $a_{\mu}^a$.

The origin the 't Hooft interaction is explained in the literature
\cite{tH} and will not be repeated here. To summarize, the low momentum
quarks are approximated by zero mode solutions in the presence of one
instanton.  Averaging over the instanton ensemble generates a vertex for
dynamical quarks, mediated by the zero modes, which can be treated
perturbatively when the instanton liquid is reasonably dilute. The
relevant small parameter is the ratio of average instanton size ($\rho$)
to the average inter-instanton spacing ($R$).  From phenomenological
\cite{ES}, variational \cite{DP1}, and lattice calculations \cite{latsize}
one finds 
\begin{equation} 
\frac{\rho}{R} \simeq \frac{1}{3}\,.
\end{equation} 
More details follow in Section 4.

We now consider the effective action itself.
In the quark sector it is a non-local one of
four-fermion operators, since quarks are connected to instantons
via the quark zero modes whose spatial extent is of the order of the
instanton size, $\rho$. Specifically, one has \cite{DD}:
\begin{equation}
S_{INT} = \lambda \int \! dU\;d^4z\;\prod_f^{N_f} \Big[ d^4x_f\;d^4y_f\;
\psi_f^{\dagger}(x_f) \dslash \Phi(x_f-z,U) \tilde\Phi(y_f-z,U) \dslash
\psi^f(y_f)\Big], 
\label{oldsint} 
\end{equation} 
where $U$ is the
instanton's $2\times N_c$ color/spin orientation matrix and $z$ is its
position. The $\Phi(x)$ is the zero mode solution for fermions in the
field of one instanton; in general it depends on the chemical potential. 
At $\mu = 0$ its Fourier transform is the form factor, 
\begin{equation} 
f(p) = 2 x \left[
I_1(x)K_0(x)-I_0(x)K_1(x)+\frac{1}{x}I_1(x)K_1(x) \right]_{x=p\rho/2}\,,
\label{ffac} 
\end{equation} such that $f(0)=1$. At $\mu\neq 0$ the form
factor is also known explicitly \cite{CD}.

Because of non-locality Eq.\,(\ref{oldsint}) is not gauge invariant.
When calculating bulk vacuum properties one needn't worry about
quantum corrections as observables are seldom sensitive to them,
however in this case the dependence is crucial.
A limited literature does exist in which non-local interactions
are modified to be gauge invariant \cite{BKN,BB,PB,WB} and we follow the
same procedure here. In particular, we will minimally modify the effective
action (\ref{oldsint}) to suit the present needs.

We will take the non-local four-fermion interaction as the starting
point. It then becomes a matter of multiplying each quark operator by a
path-ordered exponential in the background of the perturbative gluon field
$a_\mu$, replacing as: 
\begin{eqnarray} 
\psi(x) &\rightarrow& \psi(x) W(x,z), \nonumber\\ 
W(x,y) &=& {\cal P}{\rm exp} \left(
i\frac{g}{2}\int^x_y\! ds_{\mu} a^a_{\mu} \lambda^a\right), 
\end{eqnarray}
where $a^a_{\mu}$ is the perturbative field. As has been pointed out in
the cited works, the choice of path integrated over is not unique. Yet as
long as these factors transform as 
\begin{equation} 
W(x,y) \rightarrow U(x) W(x,y) U^{\dagger}(y)
\end{equation} 
by virtue of path ordering, the action will be gauge invariant. 
This remains true when, as in this
case, the interaction involves an explicit color average since overall
color is conserved. We are also concerned only with the static limit,
$q^2\rightarrow 0$, in which {\em results are independent of any
particular choice of path.}

Thus we write the modified interaction as a product over flavors,
\begin{eqnarray}
S_{INT} &=& \lambda \int dU\;d^4z\;\prod_f^{N_f} \Big[ d^4x_f\;d^4y_f\;
\nonumber\\ &&\!\!\!\!\times
\psi_f^{\dagger}(x_f) W(x_f,z) \dslash \Phi(x_f-z) \Phi^{\dagger}(y_f-z)
\dslash W(y_f,z) \psi^f(y_f)\Big], 
\label{sint} 
\end{eqnarray} 
noting
that this includes terms of all orders in $g a_{\mu}$. Calculating the
gluon polarization operator will require the linear and quadratic
contributions.

To this interaction term one must add the usual quark kinetic term,
minimally modified to preserve gauge invariance:
\begin{equation}
S_{KIN} = \int d^4 x \;
\psi^\dagger \gamma_\mu(i\partial_\mu+\frac{g}{2}\lambda^a a_\mu^a)\psi.
\label{qkin}
\end{equation}
Consequently, the color current obtained from the variation of the
action in respect to $a_\mu^a$ will have two contributions: one is the
ordinary one arising from the minimal coupling (\ref{qkin}) and the other
arising from the non-local interaction term (\ref{sint}).

\section{Diquark Condensation}

In this section some details of the color superconductor phase are
reviewed. All expressions are Euclidean and the notation generally follows
that of Ref.~\cite{CD}; the reader is referred to this reference for
details specific to this particular approach. The spontaneous breaking of
chiral symmetry was also a central concern in that work, whereas here we
are considering restored chiral symmetry. The main result of
Refs.~\cite{BR,CD,RSSV2,KP,VJ} is a competition between chiral and diquark
condensates at zero temperature and nonzero quark chemical potential. In
all cases, there is a phase transition from a low-density phase of
spontaneously broken chiral symmetry to a high-density one of color
superconductivity.

While technically involved, our previous results arise from an
effective four-quark interaction which allows for pairing of quarks as
in the BCS theory. The instanton model retains more of QCD's
features than {\it ad hoc} models, notably the anomalous breaking of
axial $U(1)$ and transmutation of dimensions, and is constructed in a
more systematic way from that underlying theory.

A prominent feature of the instanton approach used here is the
determination of the effective four-fermion
coupling constant $\lambda$. This constant
is determined by a saddle-point evaluation and proves to be nonlinearly
dependent on the background instanton density, $N/V$. A gap equation
obtained from a set of Schwinger-Dyson-Gorkov equations are solved to
first order in $\lambda$, done self-consistently to determine the quark
pairing gap, $\Delta_0$. Standard quark propagators are necessarily split
by a nonzero $\Delta_0$, since this introduces a color bias and quarks are
no longer color degenerate. This detail will be crucial when quark loops
are computed in Section 7.
\begin{figure}[tb] 
\setlength\epsfxsize{5cm}
\centerline{\epsfbox{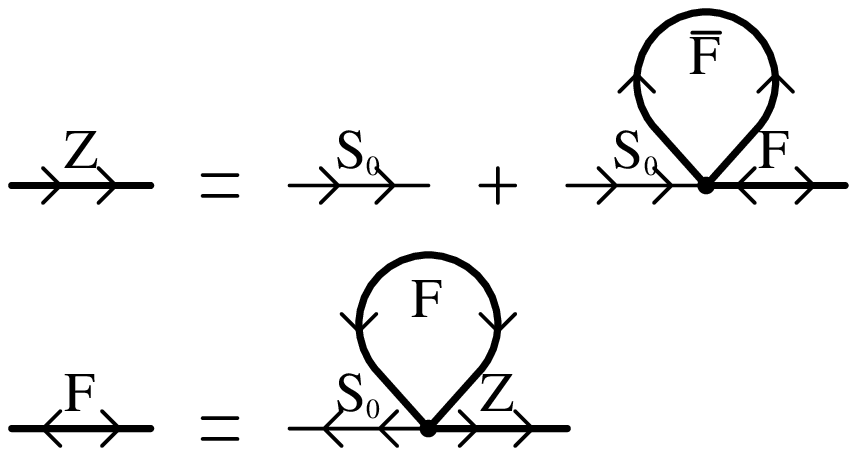}} 
\caption{Schwinger-Dyson-Gorkov diagrams for normal and anomalous 
quark propagators.  The labels refer to the functions in the text.} 
\label{sdgfig} 
\end{figure}

Since this paper deals only with a phase of chiral-symmetric, diquark
condensation the Schwinger-Dyson-Gorkov equations and corresponding
diagrams are the simple ones of Fig.~\ref{sdgfig}. We define the quark
propagators as 
\begin{equation}
\langle\psi^{f\alpha i}(p)\psi^\dagger_{g\beta j}(p)\rangle 
=  \left\{ \begin{array}{ll}
\delta^f_g \delta^\alpha_\beta S_1(p)^i_j & \quad
\alpha,\beta=1,2 \\
\delta^f_g \delta^\alpha_\beta S_2(p)^i_j & \quad
\alpha,\beta=3  \end{array}
\right. ,
\end{equation} 
and the anomalous Gorkov propagator as 
\begin{equation}
\langle\psi_L^{f\alpha i}(p)\psi_L^{g\beta j}(-p)\rangle =
\langle\psi_R^{f\alpha i}(p)\psi_R^{g\beta j}(-p)\rangle =
\epsilon^{fg}\epsilon^{\alpha\beta3}\epsilon^{ij} F(p)\,. 
\label{gorkprop}
\end{equation} 
In these expressions indices $f$ and $g$ refer to flavor,
$i$ and $j$ to spin, the Greek letters to color, and $\psi_{L,R}$ are
chiral spinors. Written in the chiral $L,R$ basis, the $4\times 4$
propagator $S_1(p)$ is of the form: 
\begin{equation} 
S_1(p) = \left[\begin{array}{cc}0 & Z(p){\bf S}_0(p)^+ \\ 
Z(p){\bf S}_0(p)^- & 0 \end{array}\right] \,, 
\end{equation} 
while $S_2(p)$ is the free
propagator for color 3 and hence identical to the above with the function
$Z(p)$ absent. The notation is $x^{\pm}=x_\mu\sigma_\mu^{\pm}$, where the
$2 \times 2$ matrices $\sigma_\mu^{\pm} = (\pm i \vec \sigma,1)$ decompose
the Dirac matrices into chiral components. The off-diagonal, bare
propagator is therefore written ${\bf S}_0(p)^{\pm} =
\left[p^\pm\right]^{-1}$.

With these definitions, we have the scalar Schwinger-Dyson-Gorkov equations
\begin{eqnarray}
Z(p) &=& 1 - \Delta(p) F(p) \nonumber\\
F(p) &=& \frac{\Delta(p) Z(p)}{p^2}\,.
\label{sdg}
\end{eqnarray}
The momentum dependence of the gap,
\begin{equation}
\Delta(p) = \Delta_0 f(p)^2 \,,
\end{equation}
is given by the instanton-induced form factor, Eq.\,(\ref{ffac}),
which suppresses the interaction beyond $p > 1/\rho \simeq
600$ MeV.
This pair of equations leads to a gap equation for $\Delta_0$,
\begin{eqnarray}
\Delta_0^2 &=& \frac{2\lambda}{N_c(N_c-1)}\int\!
\frac{d^4p}{(2\pi)^4}\: \Delta(p)F(p) \nonumber\\
&=& \frac{2\lambda}{N_c(N_c-1)}\int\!
\frac{d^4p}{(2\pi)^4}\: \frac{\Delta(p)^2}{p^2+\Delta(p)^2}\,,
\end{eqnarray}
which must be self-consistently solved with $\lambda$.
This coupling constant is in turn determined by a saddle-point
integration (exactly in the thermodynamic limit $N,V\rightarrow
\infty$),
\begin{equation}
\lambda = \frac{4 N_c (N_c-1)}{N/V} \Delta_0^2\,.
\end{equation}
Combining these two equations $\lambda$
may be eliminated and we obtain
\begin{equation}
1=8\left( \frac{N}{V} \right)^{-1} \int\! \frac{d^4p}{(2\pi)^4}\:
\frac{\Delta(p)^2}{p^2+\Delta(p)^2}\,.
\label{gapeq}
\end{equation}
The scale of the gap, therefore, is set by the instanton density $N/V$. To
this order $N/V$ remains at its vacuum value, and although it will be
affected by the finite density of quarks this is an ${\cal O} (\lambda)$
correction to the instanton weight and not considered here.

Solved numerically, the gap is $\Delta_0 \simeq 400$ MeV in vacuum and
drops to $\Delta_0 \simeq 200$ MeV at the phase transition from chiral
broken to color superconducting matter, as detailed in
Ref.~\cite{CD}\footnote{ In Ref.~\cite{CD} $\Delta_0$ was defined as half
this paper's (and the more standard) definition.}.

\section{Color Current}

\begin{figure}[bt]
\setlength\epsfxsize{4.2cm} 
\centerline{\epsfbox{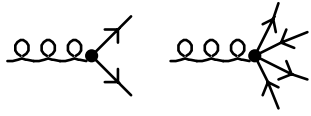}}
\caption{Quark-gluon vertices of order $g$ (left) and $g\lambda$ (right)
interactions.} 
\label{verts} 
\end{figure}

A color-conserving Noether current naturally follows from the modified
interaction, including contributions from both $S_{KIN}$ (\ref{qkin}) and
$S_{INT}$ (\ref{sint}). Along with the standard quark-gluon coupling
piece, 
\begin{equation} 
j_{\mu}^{a}(q) = -\int\! \frac{d^4p}{(2\pi)^4}
\left[ \psi_R^{\dagger}(p) \frac{\lambda^a}{2} \sigma_{\mu}^- \psi_L(p+q)
+ \psi_L^{\dagger}(p) \frac{\lambda^a}{2} \sigma_{\mu}^+ \psi_R(p+q) \right], 
\end{equation} 
the current now includes a four-quark coupling
to the gluon as shown in Fig.~\ref{verts} and written in terms of
four-momentum: 
\begin{eqnarray} 
\tilde \jmath_{\mu}^{a}(q) 
&=& \left.\frac{ \delta S_{INT} }{ \delta a_{\mu}^a(q) }
\right\vert_{a_{\mu}^a=0} \nonumber\\ 
&=& \lambda \frac{ \epsilon^{f_1f_2}\epsilon_{g_1g_2} }{4} 
\int {\cal D} \;
ie^{-ix_f\cdot(p_f-p_f')+iy_f\cdot(k_f+k_f') - iz\cdot (p_f'+k_f')} 
\nonumber\\ &&\fpj \times \Bigg[
\int_{x_1}^z\! ds_{\mu} e^{-iq\cdot s} \psi_{Lf_1}^{\dagger}(p_1)
\lambda^a {\cal M}(p_1',k_1') \psi_L^{g_1}(k_1) 
\psi_{Lf_2}^{\dagger}(p_2) {\cal M}(p_2',k_2') \psi_L^{g_2}(k_2)
\nonumber\\ &&\fpj
+ \int_z^{y_1}\! ds_{\mu} e^{-iq\cdot s} \psi_{Lf_1}^{\dagger}(p_1) 
{\cal M}(p_1',k_1') \lambda^a \psi_L^{g_1}(k_1) 
\psi_{Lf_2}^{\dagger}(p_2) {\cal M}(p_2',k_2') \psi_L^{g_2}(k_2) \Biggr] 
\nonumber\\ &&\fpj
+ (L\rightarrow R).
\end{eqnarray}
Indices on quarks denote chirality and flavor. The measure is
\begin{equation} 
{\cal D} = dU d^4z \prod_f^{N_f=2} d^4x_f d^4y_f \frac{
d^4p_f d^4p_f' d^4k_f d^4k_f' }{(2\pi)^{16}} \,
\end{equation} 
and the form factors lie in the color/spin matrices 
\begin{equation} 
{\cal M}(p,k)^{\alpha i}_{\beta j} = U^\alpha_k \epsilon^{ki} \epsilon_{jl}
U^{\dagger l}_{\beta} f(p) f(k). 
\end{equation}
The $U$ are $2\times N_c$ color orientation matrices, averaged in each
vertex. Through use of the Dirac equation for $\psi$ , one can
explicitly verify that $q_{\mu} J_{\mu}(q) = 0$ for
$J_{\mu}=j_{\mu}+\tilde \jmath_{\mu}$. While this condition would
remain satisfied with a transverse addition to the color current, no
such addition is motivated here. 

In practice we are interested in
pairing off two of the four quark legs of the second vertex. Since we are
considering a phase where chiral symmetry is unbroken, a
chirality-violating $\langle \psi^{\dagger}\psi\rangle$ loop cannot
contribute. Thus we pair either $\langle \psi \psi \rangle$ or $\langle
\psi^\dagger\psi^\dagger \rangle$ and obtain the effective current,
\begin{eqnarray} 
{\tilde \jmath}_{\mu}^a(q) &=& \frac{\Delta_0}{2} \int
d^4z d^4x \frac{ d^4p_f d^4p_f' d^4p'}{(2\pi)^{12}} 
\nonumber\\ && \times
\Big[ e^{-ix\cdot(p_1-p')-iz\cdot(p'+p_2)} 
\int_x^z ds_{\mu} e^{-iq\cdot s} 
\psi^{\dagger}_{L}(p_1) \lambda^a{\cal N}(p_2,p') \psi_{L}^{\dagger}(p_2) 
\nonumber\\ && 
+ e^{-ix\cdot(p_1+p')-iz\cdot(p'-p_2)} 
\int_z^x ds_{\mu} e^{-iq\cdot s}\psi_L(p_1) {\cal N}(p_2,p')^{\dagger}
\lambda^a \psi_{L}(p_2)\Big] 
\nonumber\\ &&
+(L\rightarrow R)\,. 
\label{newj} 
\end{eqnarray} 
where here the flavor/color/spin structure is 
\begin{equation}
{\cal N}(p,p')^{ff',\alpha\alpha',ii'} = i
\epsilon^{ff'}\epsilon^{\alpha\alpha'3} \epsilon^{ii'}f(p)f(p')\,.
\label{current} 
\end{equation} 
The two terms in Eq.\,(\ref{newj}) correspond to pairing either two 
incoming or two outgoing quark legs of Fig.~\ref{verts}.

Gluon mass terms are found in the $q^2\rightarrow 0$ limit.
When the gluon couples to nonsingular composite quark modes one can
set $q^2$ strictly to zero and the path integrals simplify to
\begin{eqnarray}
\left. e^{-ip_i'\cdot (z-x_i)} \int_{x_i}^z ds_\mu e^{-iq\cdot s}
\right\vert_{q = 0} &=& e^{-ip_i'\cdot (z-x_i)} (z-x_i)_{\mu} 
\nonumber\\ &=&
-i e^{-ip_i'\cdot (z-x_i)} \frac{\partial}{\partial p_{i\mu}'}\,.
\end{eqnarray}
This substitution, the result of which clearly depends only on the
endpoints of the path $s_{\mu}$ rather than any particular choice of
path, leads to differentiation of the form factor.
On the other hand, the
vertex can also couple gluons to Nambu-Goldstone modes which are singular
as $q^2\rightarrow 0$. Their $1/q^2$ behavior must be countered by
expanding ${\tilde \jmath}_{\mu}^a(q)$ to order $q_{\mu}$, but before this
complication arises the Nambu-Goldstone modes must be specified.

\section{Nambu-Goldstone Modes}

A symmetry has been spontaneously broken and Nambu-Goldstone modes
are certain to follow.
With $SU(3)$ being broken to $SU(2)$ there are five massless modes
and, directly analogous to the Higgs mechanism, they do not become
additional degrees of freedom.  Instead they mix with and are
incorporated into the five massive gluons, thereby relevant in the
color Meissner effect.

\begin{figure}[bt] 
\setlength\epsfxsize{10cm}
\centerline{\epsfbox{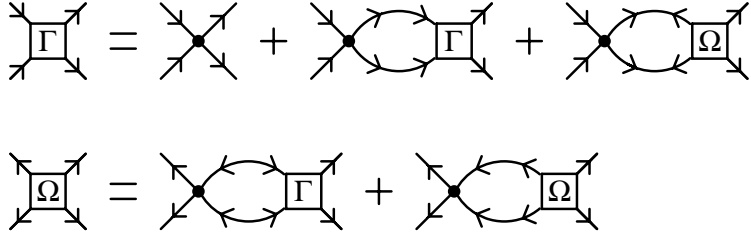}} 
\caption{Bethe-Salpeter-Gorkov diagrams for
summation of the quark-quark correlation functions.}
\label{bsgfig}
\end{figure}

Since the associated condensate is $\langle qq\rangle$,
these massless modes will be quark bilinears.
To determine their quantum numbers one need only perform a gauge rotation
on the diquark condensate and catalog the five orthogonal correlators
which appear.  Recalling the condensate direction as chosen in
Eq.\,(\ref{gorkprop}),
\begin{equation}
i\epsilon_{ij}\epsilon_{fg} \langle \psi_L^{fi\,T} \lambda^2
\psi_L^{gj}\rangle\sim\Delta_0\,,
\end{equation}
we can list the five diquark operators which couple
to the Nambu-Goldstone excitations: 
\begin{eqnarray}
&&\fpj\epsilon_{fg}\epsilon_{kl} \psi_L^{fk\,T} \lambda^7
\psi_L^{gl},\quad  i\epsilon_{fg}\epsilon_{kl} 
\psi_L^{fk\,T} \lambda^7 \psi_L^{gl},\quad
\epsilon_{fg}\epsilon_{kl} \psi_L^{fk\,T} \lambda^5
\psi_L^{gl},\nonumber\\  
&&\fpj i\epsilon_{fg}\epsilon_{kl} 
\psi_L^{fk\,T} \lambda^5 \psi_L^{gl}, \quad
\epsilon_{fg}\epsilon_{kl} \psi_L^{fk\,T} \lambda^2
\psi_L^{gl}\, .
\label{ngmodes} 
\end{eqnarray}
Propagators for these quark-quark modes will exhibit a simple pole at
$q^2=0$, behavior recovered by computing the corresponding correlation
functions. In order to obtain this pole we must sum $s$ and $u$ channel
contributions to all orders in the four-quark coupling $\lambda$
\footnote{A related summation could be done for $t$ channel diquark exchange 
as well, however these do not contribute the dominant $1/q^2$ behavior in the
limit of zero transfer momentum.}.
Since connecting these modes to the gluon propagator will be itself of order
$g^2$, we do not include any gluonic corrections to the instanton
vertex. With both standard and anomalous quark propagators at our
disposal we obtain the set of coupled Bethe-Salpeter-Gorkov equations
diagramed in Fig.~\ref{bsgfig}. The diagram shows not only standard
two-body propagators, denoted $\Gamma$, but also its anomalous
analog, $\Omega$. It is easy to see that $\Omega$ will vanish when
any of the external lines are quarks of color 3, since the vertices
conserve color and the internal (Gorkov) propagators involve only
colors 1 and 2.  

Knowing the quantum numbers of the Nambu-Goldstone modes, we can
immediately write down ans\"atze for the four-point functions which will
be required when we later compute the gluon polarization operator:
\begin{eqnarray} 
&&\fpj\langle\psi_{\chi f\alpha i}^{\dagger}(p)\psi_{\chi
f'\alpha'i'}^{\dagger}(p') \psi_{\chi'}^{g\beta
j}(k)\psi_{\chi'}^{g'\beta' j'}(k') \rangle =
-\frac{\epsilon_{ff'}\epsilon^{gg'}\epsilon_{ii'}\epsilon^{jj'}}{N_f^2(N_c
-1)} 
[\epsilon_3\lambda^a]_{\alpha\alpha'}
[\epsilon^3\lambda^b]^{\beta\beta'} 
\nonumber\\ &&\fpj\quad\qquad\qquad \times
(2\pi)^4\delta^4\left(p+p'-k-k'\right)
f(p)f(p')f(k)f(k') \Gamma_{\chi\chi'}^{ab}(p+p') \\[3mm]
&&\fpj \langle \psi_{\chi}^{f\alpha i}(p)\psi_{\chi}^{f'\alpha' i'}(p')
\psi_{\chi'}^{g\beta j}(k)\psi_{\chi'}^{g'\beta' j'}(k') \rangle 
= -\frac{\epsilon^{ff'}\epsilon^{gg'}\epsilon^{ii'}\epsilon^{jj'}}{N_f^2(N_c
-1)} 
[\epsilon_3\lambda^a]^{\alpha\alpha'}
[\epsilon_3\lambda^b]^{\beta\beta'} 
\nonumber\\ &&\fpj\quad\qquad\qquad \times
(2\pi)^4\delta^4\left(p+p'+k+k'\right)
f(p)f(p')f(k)f(k') \Omega_{\chi\chi'}^{ab}(p+p')\,. 
\end{eqnarray} 
Here,
$\left[\epsilon_3\right]_{\alpha\beta}=\epsilon_{3\alpha\beta}$, and the
$\chi$ refer to $L$ or $R$ chirality with the chiral substructures defined
as 
\begin{equation} 
\Gamma = \left[\begin{array}{cc}\Gamma_{LL} &
\Gamma_{LR} \\ \Gamma_{RL} & \Gamma_{RR} \end{array}\right] \qquad\qquad
\Omega = \left[\begin{array}{cc}\Omega_{LL} & \Omega_{LR} \\ \Omega_{RL} &
\Omega_{RR} \end{array}\right] \,, 
\end{equation} 
in which color indices
have been suppressed. It is clear from their definitions that
$\Gamma_{LR}=\Gamma_{RL}$, $\Gamma_{LL} = \Gamma_{RR}$,
$\Omega_{LR}=\Omega_{RL}$, and $\Omega_{LL} = \Omega_{RR}$.

Written in terms of these chiral elements, the two diagrams of
Fig.~\ref{bsgfig} correspond to the following set of equations:
\begin{eqnarray}
\Gamma_{LL}^{ab}(q) &=& \bar\lambda \left[ 1
+ \hat{c}_1^a\Gamma_{LR}^{ab}(q) {\cal I}_1(q)
+ \hat{c}_2^a\Gamma_{LR}^{ab}(q) {\cal I}_2(q)
+ \hat{c}_3^a\Omega_{LL}^{ab}(q) {\cal I}_3(q)
\right] \nonumber\\
\Gamma_{LR}^{ab}(q) &=& \bar\lambda \left[
\hat{c}_1^a\Gamma_{LL}^{ab}(q) {\cal I}_1(q)
+ \hat{c}_2^a\Gamma_{LL}^{ab}(q) {\cal I}_2(q)
+ \hat{c}_3^a\Omega_{LR}^{ab}(q) {\cal I}_3(q)
\right] \nonumber\\
\Omega_{LL}^{ab}(q) &=& \bar\lambda \left[
\hat{c}_3^a\Gamma_{LL}^{ab}(q) {\cal I}_3(q)
+ \hat{c}_2^a\Omega_{LR}^{ab}(q) {\cal I}_2(q)
\right] \nonumber\\
\Omega_{LR}^{ab}(q) &=& \bar\lambda \left[
\hat{c}_3^a\Gamma_{LR}^{ab}(q) {\cal I}_3(q)
+ \hat{c}_2^a\Omega_{LL}^{ab}(q) {\cal I}_2(q) \right]\,,
\label{bsgeqns}
\end{eqnarray}
where no sums are implied on color indices $a,b$.
The adjoint color vectors $\{\hat{c}_i\}$ determine the internal
propagators and are found to be
\begin{eqnarray}
\hat{c}_1 &=& (0,0,0,1,1,1,1,0) \nonumber\\
\hat{c}_2 &=& (1,1,1,0,0,0,0,1) \nonumber\\
\hat{c}_3 &=& (-1,-1,-1,0,0,0,0,1)
\end{eqnarray}
The integrals which result from these loops are:
\begin{eqnarray}
{\cal I}_1(q^2) &\equiv & \frac{1}{\Delta_0^2} \int \frac{d^4p}{(2\pi)^4}
\frac{p_+\cdot p_-}{p_+^2p_-^2} [\Delta(p_+) \Delta(p_-)]
\left[ Z(p_+)+Z(p_-) \right]
\nonumber\\
{\cal I}_2(q^2) &\equiv & \frac{2}{\Delta_0^2} 
\int \frac{d^4p}{(2\pi)^4} [\Delta(p_+) \Delta(p_-)]
F(p_+) F(p_-) \nonumber\\ 
{\cal I}_3(q^2) &\equiv & \frac{2}{\Delta_0^2} \int
\frac{d^4p}{(2\pi)^4} \frac{p_+\cdot p_-}
{p_+^2p_-^2} [\Delta(p_+) \Delta(p_-)] Z(p_+) Z(p_-) \,,
\end{eqnarray} 
where we have defined $(p_{\pm})_{\mu}=p_\mu \pm \smhalf q_\mu$
and
\begin{equation} 
\bar\lambda \equiv \frac{\lambda}{N_c(N_c-1)}\,.
\end{equation}

The set of Bethe-Salpeter-Gorkov equations (\ref{bsgeqns}) can easily be
solved for correlations functions of all adjoint colors. For colors $a=b=4
\dots 8$, in direct analogy to similar calculations of pions in the
instanton vacuum \cite{DP2}, the $q^2=0$ pole arises due to a cancellation
in the denominator which is a direct consequence of the gap equation.  The
calculation follows identically for $a = 4,5,6,$ and 7, where we find
\begin{equation} 
\Gamma_{LR}^{44}(q^2) = \Gamma_{LL}^{44}(q^2) =
\frac{\bar\lambda}{1-\bar\lambda^2{\cal I}_1(q^2)^2}\,. 
\end{equation}
After writing the gap equation (\ref{gapeq}) in these terms,
\begin{equation} 
1 = \bar\lambda {\cal I}_1(0)\,, 
\end{equation} 
we have
\begin{equation} 
\Gamma^{44}_{LR}(q^2) = \bar\lambda \left( 1 -
[\bar\lambda{\cal I}_1(0)]^2 - q^2 \bar\lambda^2 \frac{\partial}{\partial
q^2}{\cal I}_1(q^2)^2 \right)^{-1} \equiv \frac{Z_\phi}{q^2}\,,
\label{Zdef} 
\end{equation}
where
\begin{eqnarray}
Z_\phi^{-1} &=& - 2 \left. \frac{\partial {\cal I}_1(q^2)}
{\partial q^2}\right\vert_{q^2=0} \nonumber\\ &=&
\frac{1}{\Delta_0^2} \int \frac{d^4p}{(2\pi)^4}  \Bigg\{
\frac{\Delta(p)^2 - \smhalf \Delta(p) p \Delta'(p) + \smquarter p^2
\Delta'(p)^2}{p^2[p^2+\Delta(p)^2]}
\nonumber\\ && \qquad\qquad\qquad
-\frac{\smhalf\Delta(p)^2\left[\Delta(p)-p\Delta'(p)\right]^2}
{p^2[p^2+\Delta(p)^2]^2} \Bigg\}. 
\label{wfrc} 
\end{eqnarray}
The primes denote differentiation with respect to $p$. The value of
$Z_\phi$ will change with density, since not only does it depend on
$\Delta_0$ but the integrand will exhibit functional dependence on
$\mu$. In vacuum numerical evaluation gives $Z_\phi^{-1} = 8.07 \times
10^{-3}$. 

For the $\Gamma^{88}$ and $\Omega^{88}$ solving the coupled equations
becomes more involved, since all diagrams in Fig.~\ref{bsgfig} are
present. One finds 
\begin{eqnarray} 
\Gamma^{88}_{LR}(q^2) = \Omega^{88}_{LR}(q^2) 
&=& \frac{\bar\lambda}{1-\bar\lambda^2\left[ 
{\cal I}_2(q^2)+ {\cal I}_3(q^2) \right]^2}\nonumber\\ 
&=& \frac{Z_\phi}{q^2}\,, 
\end{eqnarray} 
where the second line is obtain by
making use of the Schwinger-Dyson-Gorkov equations, Eqs.\,(\ref{sdg}), in
the limit of small $q^2$. This completes the set of Nambu-Goldstone 
modes. One can
naturally compute the correlation function for the additional diquark
correlators with $a=b=1,2,3$, and find a crucial sign difference in the
combination of integrals: 
\begin{eqnarray} 
\Gamma^{11}_{LR}(q^2) = \Omega^{11}_{LR}(q^2) &=& 
\frac{\bar\lambda}{1-\bar\lambda^2\left[ 
{\cal I}_2(q^2)- {\cal I}_3(q^2) \right]^2} \nonumber\\ 
&=& \frac{1}{q^2 + m(q)^2} \,. 
\end{eqnarray} 
The mass can be determined from an ${\cal
O}(q^2)$ expansion of the denominator, however this will not be done here
as for our purposes it is enough to verify the $a=$ 1, 2, and 3
modes are indeed massive.

\section{Gluon Polarization Operator}

With the gapped quark propagators, conserved current interactions, and
composite modes defined in the previous sections, we now compute the
leading modification to the gluon polarization operator. All diagrams prove
to be color diagonal, and so we write 
\begin{equation}
\Pi_{\mu\nu}^{ab}(q^2) = \delta^{ab}\Pi_{\mu\nu}(q^2)\,. 
\label{poldiag}
\end{equation} 
Our interest is the static limit, $q^2\rightarrow 0$, which
may be considered an effective mass.

\begin{figure}[tb] 
\setlength\epsfxsize{12cm}
\centerline{\epsfbox{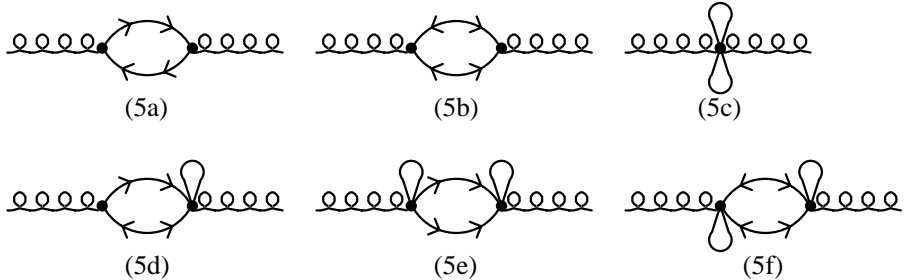}} 
\caption{Quark loop diagrams contributing to the gluon polarization operator.} 
\label{pol1} 
\end{figure}

\begin{figure}[bt] 
\setlength\epsfxsize{11cm}
\centerline{\epsfbox{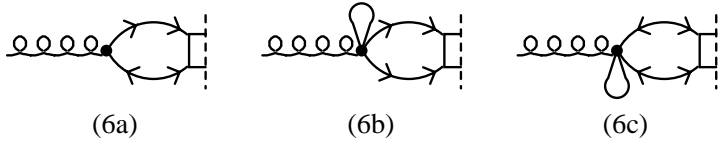}} 
\caption{Diagrams coupling the Nambu-Goldstone modes to gluons.  
Their sum squared contributes to the polarization operator.}
\label{pol2} 
\end{figure}

In the presence of a color-3, scalar diquark the gluons are
divided into three classes.
Gluons of adjoint colors 1, 2, and 3 belong to the residual $SU(2)$
gauge group and as such remain massless.
Gluons 4, 5, 6, and 7 couple one gapped quark (of fundamental color 1 or
2) with the ungapped species and share a degenerate mass. Gluon 8,
diagonal in fundamental color, obtains a mass proportional to the previous
four. One polarization operator from each class will be explained here to
avoid unnecessary repetition. All possible contributions to order $g^2$
are diagramed in Figs.~\ref{pol1} and \ref{pol2} and, depending on the
gluonic species, some of these diagrams vanish and others combine to
cancel in the static limit.

We begin with the case of gluons 4--7, considering corrections to
$\langle a_{\mu}^4 a_{\nu}^4\rangle$.
Diagrams (\ref{pol1}b), (\ref{pol1}f), and those involving (\ref{pol2}c)
require pairs of Gorkov propagators and thus vanish since the quarks of
color 3 (to which these gluons couple) cannot propagate anomalously. There
are four graphs which constitute the `contact' term proportional to
$\delta_{\mu\nu}$. 

Diagram (\ref{pol1}a) is a standard loop, where one of the quarks is
gapped (color 1 in the case of gluon 4) and the other not (color 3). After
subtracting off the vacuum part of this diagram, which remains a concern
of gluon renormalization and is not relevant to the Meissner mass, we find
\begin{equation} 
\Pi_{\mu\nu}^{(\ref{pol1}a)}(q^2\rightarrow 0) = -g^2 N_f
\delta_{\mu\nu} \int \frac{d^4p}{(2\pi)^4} \, \frac{\Delta(p)^2}{ p^2[p^2
+ \Delta(p)^2] } \,. 
\label{p4a} 
\end{equation} 
The remaining integral is
finite due to an power-law cut-off in the the function $\Delta(p)$ arising
from the finite size of instantons.

Additional diagrams arise from the modified $S_{INT}$.
Not only does this generate additional current interactions
(\ref{current}), but the interaction (\ref{sint}) itself contains a
contact contribution to the gluon two-point function. This, diagram
(\ref{pol1}c), is the second variation of the action (\ref{sint}) with
respect to the fourth gluon field, 
\begin{equation}
\Pi_{\mu\nu}^{(\ref{pol1}c)}(q^2\rightarrow 0) = \left.\frac{\delta^2
S_{INT}}{ \delta a_{\mu}^4(q) \delta a_{\nu}^4(q) } \right\vert_{q^2 =
0}\,.  
\end{equation}
Evaluated to order $g^2$ it is
\begin{equation}
\Pi_{\mu\nu}^{(\ref{pol1}c)}(q^2\rightarrow 0) = \frac{1}{4} g^2 N_f
\delta_{\mu\nu} \int \frac{d^4p}{(2\pi)^4} \, \frac{ -\smhalf p^2
\Delta'(p)^2 +\Delta(p)p^2 \Delta''(p)} {p^2[ p^2 + \Delta(p)^2]} \,.
\label{p4f} 
\end{equation} 
Diagrams (\ref{pol1}d) and (\ref{pol1}e),
constructed with the additional current piece, are 
\begin{eqnarray}
\Pi_{\mu\nu}^{(\ref{pol1}d)}(q^2\rightarrow 0) &=& \frac{1}{4} g^2 N_f
\delta_{\mu\nu} \int \frac{d^4p}{(2\pi)^4} \, \frac{ \Delta(p) p
\Delta'(p) } {p^2[ p^2 + \Delta(p)^2 ]} \nonumber\\
\Pi_{\mu\nu}^{(\ref{pol1}e)}(q^2\rightarrow 0) &=& \frac{1}{8} g^2 N_f
\delta_{\mu\nu} \int \frac{d^4p}{(2\pi)^4} \, \frac{ p^2 \Delta'(p)^2 }
{p^2 [p^2 + \Delta(p)^2] } \,. 
\label{p4de} 
\end{eqnarray}

Eqs. (\ref{p4a}), (\ref{p4f}), and (\ref{p4de}) comprise the microscopic
equivalent of the Higgs contact term (\ref{Pib}), and their sum is
\begin{eqnarray} 
&&\fpj\Pi_{\mu\nu}^{(\ref{pol1})}(q^2\rightarrow 0) = 
\nonumber\\ &&\fpj\quad - g^2 N_f
\delta_{\mu\nu} \int \frac{d^4p}{(2\pi)^4} \, \frac{ \Delta(p)^2 -
\smquarter\Delta(p) p \Delta'(p) - \smquarter \Delta(p) p^2 \Delta''(p)^2
} {p^2[ p^2 + \Delta(p)^2]} \,. 
\label{p4} 
\end{eqnarray} 
The integral can be trivially rewritten and then manipulated to yield 
\begin{eqnarray} 
&&\fpj\int \frac{d^4p}{(2\pi)^4} \, \Bigg\{\frac{ \Delta(p)^2 - \smhalf \Delta(p) p
\Delta'(p) }{p^2[ p^2 + \Delta(p)^2]}
\nonumber\\ &&\fpj \qquad \qquad +\frac{ \Delta(p) p \Delta'(p)
}{4p^2[ p^2 + \Delta(p)^2]} -\frac{ \Delta(p) p^2 \Delta''(p) }{4p^2[ p^2
+ \Delta(p)^2]}  \Bigg\} = Z_{\phi}^{-1}\Delta_0^2 \,. 
\label{ibp}
\end{eqnarray} 
The final equality is achieved by integrating the second
and third terms by parts.

The Nambu-Goldstone modes couple to the gluons as in Fig.~\ref{pol2}. The
construction
[(\ref{pol2}a)+(\ref{pol2}b)]$\Gamma^{44}(q^2)$
[(\ref{pol2}a)+(\ref{pol2}b)]
supplies the $q_{\mu}q_{\nu}/q^2$ piece to ensure transversality:
\begin{eqnarray} 
\Pi_{\mu\nu}^{(\ref{pol2})}(q^2\rightarrow 0) &=& g^2
N_f q_{\mu}q_{\nu} \frac{Z_\phi}{q^2} 
\Bigg[\int \frac{d^4p}{(2\pi)^4}\, \Bigg\{
\frac{\smhalf\Delta(p)^2\left[\Delta(p)-p\Delta'(p)\right]^2}
{p^2\left[p^2+\Delta(p)^2\right]^2} 
\nonumber\\ &&\qquad
- \frac{\Delta(p)^2 - \smhalf \Delta(p) p \Delta'(p) + \smquarter p^2
\Delta'(p)^2}{p^2\left[p^2+\Delta(p)^2\right]} 
\Bigg\} \Bigg]^2 \nonumber\\ 
&=& g^2 N_f \Delta_0^2 Z_\phi^{-1} \frac{q_{\mu}q_{\nu}}{q^2} \,. 
\label{p5}
\end{eqnarray} 
We have now accounted for all contributions to gluons 4, 5,
6, and 7. 

Analysis of the eighth gluon follows in a similar fashion, although with
terms from every diagram. A superficial difference lies in the factors
arising from the elements $\lambda^8$ which lead to a polarization $4/3$
times the previous result. More subtle is the combination of diagrams
(\ref{pol1}a) and (\ref{pol1}b). For $a_{\mu}^8$ this sums to $4/3$ times
Eqs.\,(\ref{p4a}), whereas for $a_{\mu}^{1,2,3}$ they cancel one another.
This cancellation is only manifest to ${\cal O}(\Delta_0^4)$ here, due to
the fact that the diagrams are constructed with a pair of one-loop,
resummed quark propagators as determined in Section 3. This generates a
$\Delta(p)^2$ term in each integrand denominator, in essence including
higher-order terms in the perturbative expansion (in $\lambda$) which
violate gauge invariance. After the vacuum pieces are subtracted, these
diagrams contribute the following to $a_{\mu}^8$: 
\begin{eqnarray}
\Pi_{\mu\nu}^{(\ref{pol1}a+\ref{pol1}b)}(q^2\rightarrow 0) 
&=& \frac{1}{3} g^2 N_f \Delta_0^2 \delta_{\mu\nu} \int \frac{d^4p}{(2\pi)^4} 
\Bigg\{ \frac{p^2}{[p^2+\Delta(p)^2]^2}
\nonumber\\ && \qquad\qquad 
- \frac{2\Delta(p)^2}{[p^2+\Delta(p)^2]^2} 
- \frac{1}{p^2}\Bigg\} \nonumber\\ 
&=&
-\frac{4}{3} g^2 N_f \Delta_0^2 \delta_{\mu\nu} \int \frac{d^4p}{(2\pi)^4}
\Bigg\{ \frac{\Delta(p)^2}{p^2[p^2+\Delta(p)^2]} 
+ {\cal O}(\Delta_0^4)\Bigg\} \,. \nonumber\\
\end{eqnarray} 
For gluons 1, 2, and 3, the sign of the second
term is changed and the corresponding integral is of order $\Delta_0^4$,
and thus this quantity vanishes to the order to which we know quark
propagators, $\Delta_0^2$ . The same follows for pairs of diagrams with
similar construction, such as (\ref{pol1}e) and (\ref{pol1}f).

By combining Eqs. (\ref{poldiag}), (\ref{p4}), (\ref{ibp}), and (\ref{p5})
we arrive at the satisfyingly compact expressions 
\begin{equation}
\Pi_{\mu\nu}^{ab}(q^2\rightarrow 0) =
\left\{ \begin{array}{ll} 0 
&\, a,b=1,2,3 \\ -g^2 N_f \Delta_0^2 Z_\phi^{-1} \delta^{ab} 
\left( \delta_{\mu\nu} - \frac{q_{\mu}q_{\nu}}{q^2} \right) 
&\, a,b = 4,5,6,7 \\ 
-\frac{4}{3}g^2N_f\Delta_0^2 Z_\phi^{-1} \delta^{ab} 
\left( \delta_{\mu\nu} - \frac{q_{\mu}q_{\nu}}{q^2} \right) 
&\, a,b = 8 .\end{array} \right. 
\label{Pifin} 
\end{equation} 
Transversality requires
that the contact term be proportional to the wave function renormalization
of the Nambu-Goldstone modes and this result, a Ward Identity for color
superconductivity, is recovered here.

Finally, to determine a numerical value for the masses we must fix the
coupling constant $g$.  Evaluating in the instanton vacuum, one finds the
large finite action \cite{DP1} 
\begin{equation} S_0 = \frac{8 \pi^2}{g^2} \simeq 12 \,, 
\end{equation} 
or $g \simeq 2.6$ and the perturbative
expansion parameter 
$\alpha_s = g^2/4\pi = 0.54$. 

The gluon masses squared are thus
\begin{equation}
M_a^2 = \left\{\begin{array}{ll}
 0 &\qquad a=1,2,3 \\
2 g^2 \Delta_0^2 Z_\phi^{-1}
\simeq (150 \, {\rm MeV})^2 &\qquad a=4,5,6,7 \\
\frac{8}{3} g^2 \Delta_0^2 Z_\phi^{-1}
\simeq (175 \, {\rm MeV})^2 &\qquad a=8.\end{array} \right.
\label{massesnum}
\end{equation}
These masses apply to the vacuum, $\mu = 0$.

\begin{figure}[bt]
\setlength\epsfxsize{11cm} 
\centerline{\epsfbox{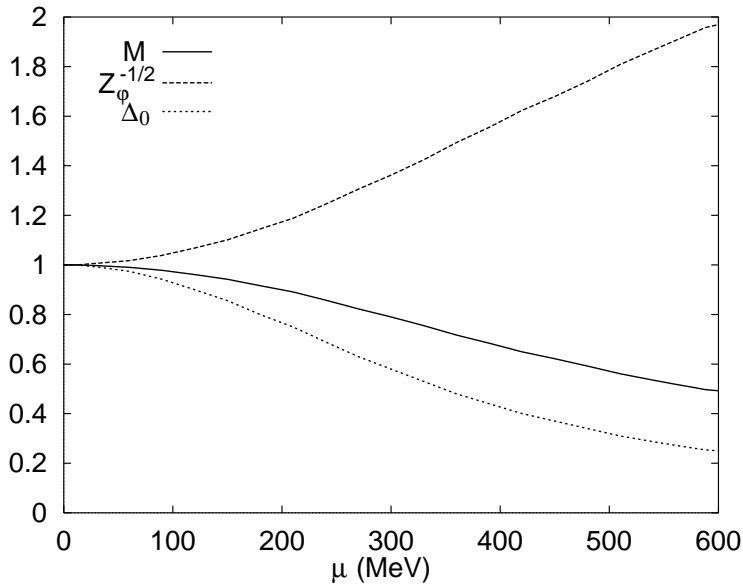}}
\caption{Scaling of $M$, $Z_\phi^{-1}$, and $\Delta_0$ with quark chemical
potential $\mu$.  Each quantity is shown relative to its vacuum value. For
reference, diquark condensation is only stable in quark matter at and
above $\mu \simeq 300$ MeV.} 
\label{finited} 
\end{figure}

In order to estimate the finite-density behavior of the Meissner mass, we
can simply take the values of $\Delta_0$ for a given $\mu$ from the
results of Ref.~\cite{CD}. As detailed in that paper, the instanton form
factor (\ref{ffac}) becomes density dependent and thus $\Delta(p)$ should
be replaced by a complicated $\Delta(p_4, |\vec{p}|,\mu)$. However, the
changes in $Z_\phi$ arising from the finite-$\mu$ modifications of the
form factors are minor compared compared to the changes in the gap
magnitude, $\Delta_0$; we numerically estimate this correction to be about
3\%. For simplicity we therefore considered only the scaling from
$\Delta_0(\mu)$ (taken from previous work \cite{CD}). 
Each non-zero gluon mass is proportional to $\Delta_0 Z_\phi^{-1/2}$ (see 
Eq.\,(\ref{massesnum})) and therefore all will scale identically with density.
The resulting gluon mass $M(\mu)$,
as well as that of the renormalization constant
$Z_\phi(\mu)^{-1/2}$, is shown in units of its vacuum value in
Fig.~\ref{finited}. Note that although $Z_\phi$ rises with increasing
quark density, this effect is not sufficient to overcome the falling gap
and the masses continuously decrease. The first point of
physical relevance would occur around $\mu \simeq 300$ MeV, the common
prediction for chiral restoration to a color superconductor. Results
for matter at lower densities correspond to an unstable solution and
are only of academic interest \cite{CD,BR,RSSV2,KP,VJ}. 

\section{Similarity with Chiral Symmetry Breaking}

Apart from the gauge coupling $g$ the gluon masses (\ref{massesnum})
are determined by the combination $F_{qq}^2\equiv 2\Delta_0^2Z_\phi^{-1}$.
This quantity is the analog of the $F_\pi^2$ constant in the chiral-broken
phase, both in its physical meaning and algebraically.

In the chiral-broken phase the Nambu-Goldstone bosons are pions. The
correlation function of the axial current,
$j_{\mu 5}^A=\psi^{\dagger}\gamma_\mu\gamma_5\tau^A\psi$, in the
massless quark limit has the transverse form
\begin{equation}
\langle j_{\mu 5}^A(q)j_{\nu 5}^B(-q)\rangle
=\Pi_{\mu\nu}^{AB}(q)=-F_\pi^2\delta^{AB}\left(\delta_{\mu\nu}
-\frac{q_\mu q_\nu}{q^2}\right)\,.
\label{Piaxial}\end{equation}
The transversality of $\Pi_{\mu\nu}$ is the consequence of the
conservation of the axial current; the $1/q^2$ pole is due
to the pion in the intermediate state. Were there gauge bosons
coupled to the quark axial current their mass would be equal
to $F_\pi$ multiplied by the corresponding gauge coupling. In case
of diquark (vs. quark-antiquark) condensation the relevant currents are
color ones, and we obtain a similar form for the correlation function of
two color currents, Eq.~(\ref{Pifin}). One needs only to multiply
$F_{qq}$ by the gauge coupling to deduce the gluon mass.

If instantons are dilute, the leading contribution to the Kronecker
part of the polarization operator arises from the $\Pi_{\mu\nu}^{(4a)}$
piece, Eq.~(\ref{p4a}). It is the only contribution which diverges
logarithmically if one neglects the momentum dependence of the gap
$\Delta(p)$. For the same reason the leading contribution to the
$Z_\phi^{-1}$ and $F_{qq}^2$ in the dilute limit comes from the pieces not
containing the derivatives $\Delta^\prime(p)$. One has therefore in the
dilute limit:
\begin{equation}
F_{qq}^2\simeq 2\int\!\frac{d^4p}{(2\pi)^4}\frac{\Delta(p)^2}
{p^2[p^2+\Delta(p)^2]} \simeq
\frac{2\Delta_0^2}{8\pi^2}\log\frac{R^2}{\rho^2}
\label{Fqqapp}\end{equation}
where $\Delta_0$ is the superconducting gap
at zero momentum. It follows from the gap equation (\ref{gapeq}) that
$\Delta_0\sim \pi\rho\sqrt{N/V} =\pi\rho/R^2$. Similarly, the axial
correlation function (\ref{Piaxial}) computed in Ref.~\cite{DP2}
gives (in the same approximation)
\begin{equation}
F_\pi^2\simeq 4N_c\int\!\frac{d^4p}{(2\pi)^4}\frac{M(p)^2}{[p^2+M(p)^2]^2}
\simeq \frac{4N_cM_0^2}{8\pi^2}\log\frac{R^2}{\rho^2}
\label{Fpiapp}\end{equation}
where $M(p)$ is the dynamical quark mass (the chiral gap) whose value
at zero momentum is determined from a corresponding gap equation to be
$M_0\sim \pi\rho\sqrt{N/VN_c}$.

We see, thus, that not only have the Meissner mass and the $F_\pi$
constant analogous meaning, but their algebraic structure is quite
similar. One expects, therefore, that the numerical value for the
Meissner mass is of the order of $F_\pi$, and this expectation is
confirmed by an exact numerical calculation of the previous section.

\section{Discussion and Conclusions}

We have analyzed the problem of spontaneous gauge symmetry breaking
brought about by a diquark condensate.
Since the broken symmetry is continuous and gauged the resulting
Nambu-Goldstone modes do not remain in the spectra, instead mixing
with the longitudinal components of the gauge fields to produce massive
gauge bosons. This, a dynamical Higgs mechanism, can be called the color
Meissner effect in the context of color superconductivity.

To reveal the gauge boson masses mathematically one has to compute
the polarization operator, which ought to be transverse,
\begin{equation}
\Pi_{\mu\nu}^{ab}(q)=-M^2\delta^{ab}\left(\delta_{\mu\nu}
-\frac{q_\mu q_\nu}{q^2}\right),
\label{polop}
\end{equation}
where the massless pole $1/q^2$ is the manifestation of the
Nambu--Goldstone intermediate state. The coefficient, $M^2$ gives the mass
of the gauge boson. In this paper, we explicitly solved this problem for
the case of diquark condensation as induced by the instanton background,
the effective action from which we found necessary to modify
in order to maintain a conserved color current.

Through computing the gluon polarization operators to order $g^2$
we find the effective gluon masses to be on the order of the diquark
gap. The three gluons comprising the residual $SU(2)$ group
remain massless and hence a quark-gluon medium would become
color-biased in such a phase. The analysis here was done in the limit of
zero temperature and, initially, vanishing chemical potential $\mu$,
though strictly speaking at zero $\mu$ the color superconductor is only
metastable with the ground state being the usual chiral-broken phase.  We
then estimated finite-density dependence of the calculated quantities. At
the critical density, where one expects the phase transition to the color
superconducting phase, we deduce
that the Meissner masses of the $4^{th}$, $5^{th}$, $6^{th}$,
and $7^{th}$ gluon are about 120 MeV and the $8^{th}$ gluon has a mass
about 140 MeV. These quantities would be of physical relevance should a
low temperature, high density region become experimentally accessible. At
a chemical potential low enough to leave the instanton background
approximately unchanged ($\mu <$ 0.6 GeV), the instanton effects analyzed
here would still be present and likely dominant.

In computing the coupling of the Nambu-Goldstone modes to gluons
we have established that their mixing is described by
an effective theory in which the
composite diquark is replaced by a complex scalar in the fundamental
representation of the gauge group, $\phi^a$.  The effective Lagrangian
coincides with that of elementary scalar field covariantly coupled to
gluons, aside from an overall factor $Z_\phi$ which is the `wave function
renormalization' of the composite scalar field. There are no {\em a
priori} reasons for this factor to be close to unity (as it is in the case
of a weakly-coupled elementary Higgs field), and indeed we find a
substantial deviation from unity. Meanwhile, it is a crucial factor for
the estimate of the Meissner mass.

If electromagnetic interactions were taken into account, the gluon
$a_{\mu}^8$ will invariably mix with the (massless) photon.
This mixing, estimated from general arguments to be rather small
\cite{ABR2}, reorganizes the fields into a massive `new gluon' and
massless `new photon'. Given that the gluonic sector alone can be
recast as a Yang-Mills-Higgs theory, this additional result would
complete the analogy to the electroweak sector of the Standard Model.

Finally, we would like to comment on the phase transition between
ordinary chiral-broken phase and the color superconductivity. The
point is that all estimates existing in the literature
(see Ref.~\cite{CD} and references therein)
indicate that this phase transition happens alarmingly `early': taken
literary, the claim is that the interior region of a heavy nucleus is
actually in a `boiling' state. However, those estimates generally neglect
the influence of the dense medium on the gluonic background fluctuations
which induce the color superconductivity itself. The Meissner mass of
about 150 MeV found here is a large quantity and, together with the Debye
mass and other effects, will suppress instantons. Therefore, by taking
into account back-influence effects one has a chance to `save' ordinary
nuclear matter from a premature phase transition by moving that transition
to higher densities.

Although the instanton density is not expected to change
significantly for any chemical potential below the inverse instanton 
size (600 MeV), at and beyond this point perturbative Meissner and 
screening masses will become increasingly important. At asymptotically 
large densities perturbative gluon exchange is the source of
diquark condensation \cite{BL,SS1,Son,RSWZ,DR}, whereas in this work 
we have been concerned with the low density, nonperturbative regime. 
Determining the behavior at a moderate density scale -- which would be 
an interpolation between the two -- is necessary if one wishes to 
confidently consider signals of color superconductivity in any 
potentially realizable situation, be it in some future heavy-ion 
collider or a neutron star.

\section*{Acknowledgments}

G.W.C. thanks D. Rischke for discussions and both the Leon 
Rosenfeld Fund and USDOE grant DE-FG02-88ER40388 for support.

\end{document}